\documentclass[11pt]{article}
\usepackage{cite}
\usepackage{amsmath,amsfonts,amssymb}
\usepackage[small,bf,hang]{caption}
\usepackage{slashed}
\usepackage{mathabx}
\usepackage{latexsym,epsfig}
\usepackage{color}

\def\hybrid{
        \topmargin -20pt
        \oddsidemargin 0pt
        \headheight 0pt \headsep 0pt
        \textwidth 6.25in 
        \textheight 9.5in 
        \marginparwidth .875in
        \parskip 5pt plus 1pt \jot = 1.5ex}

\hybrid

 \csname
@addtoreset\endcsname{equation}{section}

\def\moth{\mathsurround=0pt}
\newdimen\zo \zo=0pt

\def\tick{\leaders\hrule height 0.5ex depth 0pt \hskip 0.5pt}
\def\upboxfill{$\moth \setbox\zo\hbox{\tick}%
  \hskip 3pt\hbox to 0pt{$\tick$\hss}\hrulefill \hbox to 7.5pt{$\tick$\hss}$}

\def\dtick{\leaders\hrule height .34pt depth 0.5ex \hskip 0.5pt}
\def\downboxfill{$\moth \setbox\zo\hbox{\dtick}%
  \hskip 2pt\hbox to 0pt{$\dtick$\hss}\hrulefill \hbox to 2pt{$\dtick$\hss}$}

\def\bec{\begin{center}}
\def\ec{\end{center}}
\def\a{\alpha}  

\def\b{\beta}  
 
\def\C{\Gamma}
\def\d{\delta} 

\def\e{\epsilon}

\def\om{\omega}
\def\hom{{\hat\omega}}
\def\hF{{\hat F}}
\def\bpsi{{\bar \psi}}
\def\oom{{\stackrel{\circ}{\omega}}}

\def\cL{{\cal L}}

\def\q{\quad}

\def\nn{\nonumber}

\def\be{\begin{equation}}
\def\ee{\end{equation}}
\def\bea{\begin{eqnarray}}
\def\eea{\end{eqnarray}}
\def\ba{\begin{array}}
\def\ea{\end{array}}

\begin{document}

\vskip .6cm
{\setlength{\parindent}{0cm}
{\Large \bf {Corrigendum to "Coupling the SO(2) supergravity through dimensional reduction"} 
[Phys. Lett. B 96 (1-2) (1980) 89-93]}\\
\emph{A. Chamseddine $^{1,2}$, H. Nicolai $^3$}\\
\emph{$^{1}$Physics Department, American University of Beirut, Lebanon}\\
\emph{$^2$ {I.H.E.S. F-91440 Bures-sur-Yvette, France}}\\
\emph{$^3${Max-Planck-Institut fuer Gravitationsphysik (Albert-Einstein-Institute)
		 Muehlenberg 1, D-14476, Potsdam Germany}}\\
\emph{ Emails chams@aub.edu.lb and hermann.nicolai@aei.mpg.de}\\}
\vskip .6cm
Due  to renewed interest in $D=5$ supergravity and its quartic fermionic terms, 
fuelled by very recent work on supersymmetric quantum cosmology \cite{DS},
we would like to correct an error in eq.(4) of the above paper \cite{CN} (to which
we also refer for our notations and further explanations\footnote{In particular,
  our metric signature is $(+----)$, and thus $\bar\psi_M \equiv \psi_M^\dagger\Gamma^0$.
  Furthermore, $\Gamma^{ABCDE}_{\a\b} = \epsilon^{ABCDE} \, \d_{\a\b}$ with 
  $\epsilon^{01234} = \epsilon_{01234} = 1$.}). In 1.5 order formalism 
\cite{CW} the  correct Lagrangian of pure $D=5$ supergravity reads 
\bea\label{L1} 
\cL &=& - \frac14 e R(\om) 
\,-\, \frac{i}2 e\left[\bpsi_M \C^{MNP} \overrightarrow{D}_N\left(\frac{\om + \hom}2\right) \psi_P -
   \bpsi_M  \overleftarrow{D}_N\left(\frac{\om + \hom}2\right) \C^{MNP} \psi_P\right] \\[1mm]
   && \hspace{-5mm} - \, \frac14 eF^{MN} F_{MN} 
   \,-\, \frac{\sqrt{3}}8 ie \bpsi_M X^{MNPQ} \psi_N (F_{PQ} + \hF_{PQ} )
   \,-\, \frac1{6\sqrt{3}} \e^{MNPQR} F_{MN} F_{PQ} A_R   \nn
\eea
with the 4-component Dirac vector spinor $\psi_M$.
The $(3\om - \hom)/2$ in eq.(4) of \cite{CN} must thus be replaced by $(\om + \hom)/2$. 
The terms with $\hom$ and $\hF$ account for all higher order fermionic terms
in 1.5 order formalism, as they do in $D=11$ supergravity \cite{CJS}.
Here the spin connection $\om$ is determined by its equation of motion 
\be
\om_{MAB} = \oom_{MAB}(e) \,+\, \kappa_{MAB}
\ee
with the contorsion tensor
\bea 
\kappa_{MAB} &=& - \, \frac{i}2 \bpsi_Q \C^{QR}{}_{MAB} \psi_R  \, + 
             \frac{i}2\big( \bpsi_M\C_B\psi_A - \bpsi_A\C_B\psi_M) \, + \nn\\[2mm]
&&         + \, \frac{i}2\big( \bpsi_B\C_M\psi_A - \bpsi_A\C_M\psi_B)
               \,- \, \frac{i}2\big( \bpsi_M\C_A\psi_B - \bpsi_B\C_A\psi_M)
\eea
The supercovariant spin connection and field strength are
\be
\hom_{MAB}  \equiv \om_{MAB} + \frac{i}2 \bpsi_Q \C^{QR}{}_{MAB} \psi_R \; ,\q
\hF_{MN} \equiv F_{MN} + \frac{\sqrt{3}i}2(\bpsi_M\psi_N - \bpsi_N \psi_M)
\ee
From (\ref{L1}) one obtains the supercovariant Rarita-Schwinger equation ($\equiv$ eq. (14) of \cite{CN})
\be
\C^{MNP} D_N(\hom)\psi_P + \frac{\sqrt{3}}4 X^{MNPQ}\psi_N \hF_{PQ} = 0
\ee

The Lagrangian in second order formalism, which was not explicitly written out
in \cite{CN}, is obtained in the usual way by substituting the second order spin 
connection into the Lagrangian (\ref{L1}), with the result
\bea 
\cL &=& - \frac14 e R(\oom(e)) 
\,-\, \frac{i}2 e \left[\bpsi_M \C^{MNP} \overrightarrow{D}_N\left(\oom(e)\right) \psi_P -
   \bpsi_M  \overleftarrow{D}_N\left(\oom(e)\right) \C^{MNP} \psi_P\right] \\[1mm]
   && \hspace{-5mm}
   - \, \frac14 eF^{MN} F_{MN} 
   \,-\, \frac{\sqrt{3}}4 ie \bpsi_M X^{MNPQ} \psi_N F_{PQ} 
   \,-\, \frac1{6\sqrt{3}} \e^{MNPQR} F_{MN} F_{PQ} A_R   \,+\, \cL_{\rm quartic} \nn
\eea
where
\begin{align*}
\mathcal{L}_{\mathrm{quartic}} \, & \mathcal{=}\,\frac{1}{4}e \,  \bigg[  \left(
\overline{\psi}_{M}\Gamma^{N}\psi_{N}-\overline{\psi}_{N}\Gamma^{N}\psi
_{M}\right)  \left(  \overline{\psi}^{M}\Gamma^{P}\psi_{P}-\overline{\psi}%
_{P}\Gamma^{P}\psi^{M}\right) \\
& \quad\qquad  -\frac{1}{4}\left(  \overline{\psi}_{M}\Gamma_{N}\psi_{P}-\overline{\psi
}_{P}\Gamma_{N}\psi_{M}\right)  \left(  \overline{\psi}^{M}\Gamma^{N}\psi
^{P}-\overline{\psi}^{P}\Gamma^{N}\psi^{M}\right)  \\
& \quad\qquad -\frac{1}{2}\left(  \overline{\psi}_{M}\Gamma_{N}\psi_{P}-\overline{\psi
}_{P}\Gamma_{N}\psi_{M}\right)  \left(  \overline{\psi}^{M}\Gamma^{P}\psi
^{N}-\overline{\psi}^{N}\Gamma^{P}\psi^{M}\right)  \\
& \quad\qquad \left.  +\, \overline{\psi}_{M}\Gamma^{MNPQ}\psi_{N}\overline{\psi}_{P}\psi
_{Q}+\frac{3}{2}\left(  \overline{\psi}_{M}\psi_{N}-\overline{\psi}_{N}%
\psi_{M}\right)  \overline{\psi}^{M}\psi^{N}\right]
\end{align*}
This can now be compared to \cite{Cremmer} (which, however, uses symplectic
Majorana spinors rather than Dirac spinors).\\[2mm]
{\bf Acknowledgments:} We would like to thank T.~Damour, P.~Spindel and A. Van Proeyen for 
discussions and correspondence.


\begin{thebibliography}{99}
\bibitem{DS} T.~Damour and P.~Spindel,
   Phys.Rev. {\bf D90} (2014) 103509;
   Phys.Rev. {\bf D95} (2017) 126011
\bibitem{CN} A.~Chamseddine and H.~Nicolai, Phys. Lett. {\bf 96B} (1980) 89
\bibitem{CW} A.~Chamseddine and P.~West, Nucl. Phys. {\bf B129} (1977) 39;\\
      P.K.~Townsend and P. van Nieuwenhuizen, Phys. Lett. {\bf B67} (1977) 439
\bibitem{CJS} E.~Cremmer, B.~Julia and J.~Scherk, Phys. Lett. {\bf 76B} (1978) 409
\bibitem{Cremmer} E.~Cremmer, {\it Supergravities in 5 Dimensions}, 
published in A. Salam, E. Sezgin (eds.): Supergravities in diverse dimensions, 
Cambridge University Press (1980) 
\end{thebibliography}
\end{document}